# Label-Free Detection of Governance Evidence Degradation in Risk Decision Systems

Oleg Solozobov[1*]

[1] Independent Researcher (Global)

[*] Correspondence: Oleg Solozobov (dev404ai@gmail.com)

ORCID: https://orcid.org/0009-0009-0105-7459

## Abstract

Risk decision systems in fraud detection and credit scoring operate under structural label absence: ground truth arrives weeks to months after decisions are made. During this blind period, model performance may degrade silently, eroding the governance evidence that justifies automated decisions. Existing drift detection methods either require labels (supervised detectors) or detect statistical change without distinguishing harmful degradation from benign distributional evolution (unsupervised detectors). No existing framework integrates drift detection with governance evidence assessment and operational response.

This paper presents a label-free governance monitoring extension to the Governance Drift Toolkit toolkit that produces governance alerts rather than statistical alarms. The monitoring architecture applies composite multi-proxy monitoring across four proxy monitors (score distribution, feature drift, prediction entropy, confidence distribution), with governance-calibrated thresholds.

Empirical evaluation on the Lending Club credit scoring dataset (1.37M loans, 11 years) demonstrates three findings. First, raw proxy metrics (Feature PSI delta up to 1.84, Score PSI delta up to 0.92) distinguish injected covariate degradation from natural temporal drift in an offline evaluation setting. Second, pure concept drift in $P(Y|X)$ produces exactly zero delta across all proxy metrics in all windows, confirming the irreducible blind spot of label-free monitoring as a structural verification. Third, the composite score provides monotonic severity progression as more monitors trigger (0.583 to 0.833 to 1.000), enabling graduated governance response. Cross-domain comparison with IEEE-CIS fraud detection results (Solozobov, 2026e) shows the *detectable/undetectable* boundary is consistent across both domains. The toolkit and evaluation code are available as open-source artifacts.

**Keywords:** label-free monitoring, model degradation detection, governance evidence, silent model failure, concept drift, structural detectability

## 1. Introduction

Risk decision systems in financial services – fraud detection, credit scoring, transaction monitoring – operate under a structural constraint that distinguishes them from most machine learning deployment contexts: ground truth labels arrive weeks to months after the decision. A fraud detection model makes a real-time *accept/reject* decision, but whether that decision was correct may not be known for 30 to 180 days, when dispute resolution, chargeback processes, or investigation outcomes finally produce a label. During this blind period, the model operates without feedback, and its governance evidence – the operational proof that the system is performing as intended – silently degrades if the underlying data distribution shifts.

This is not a hypothetical concern. Concept drift in production ML systems is well-documented, and the detection methods community has produced hundreds of papers addressing it. Yet the operational question facing every fraud team – "is our model still working?" – remains surprisingly difficult to answer without labels. Supervised drift detectors (DDM, EDDM, ADWIN) require immediate label availability and are inapplicable under structural delay. Unsupervised methods (PSI, KS tests, KL divergence) detect distributional change in the input space but cannot distinguish benign covariate shift from harmful concept drift that degrades model performance. Prinster et al. emphasize the growing recognition for the need to continuously monitor deployed AI systems for determining when model updates are required to mitigate downstream harm (Prinster et al., 2025).

Industry has responded with bespoke solutions. These approaches demonstrate the universal demand for label-free monitoring but remain system-specific, without governance framing or transferable architecture. The closest academic prior art – PRODEM (*Carcillo/Dal* Pozzolo) – proposes a meta-model for fraud degradation detection but uses proprietary data and is not positioned as a governance monitoring instrument.

This paper argues that governance evidence degradation in real-time risk decision systems can be reliably detected without ground truth labels through composite statistical proxy monitoring, but with characterized blind spots that define the boundary of label-free governance assurance. The contribution is not a new drift detection algorithm – the community has many – but an operational integration: combining multiple proxy signal categories into a governance monitoring architecture that produces actionable governance alerts rather than statistical alarms.

The work builds on three foundations. The governance evidence framework (Solozobov, 2026b) distinguishes governance evidence from compliance documentation, establishing that evidence quality degrades structurally over time. The evidence sufficiency model (Solozobov, 2026e) defines measurable thresholds below which governance evidence no longer supports defensible decision-making. This paper contributes the monitoring layer: continuous detection of when evidence quality crosses those thresholds, without requiring the labels that remain structurally unavailable.

Three research questions guide the investigation:

1. Can governance evidence degradation in real-time risk decision systems be reliably detected without ground truth labels, using statistical proxy monitoring?
2. What operational monitoring architecture enables continuous governance assessment under structural label absence (1-6 month blind periods)?
3. What are the structural limits of label-free monitoring, and which degradation types remain undetectable across domains?

To answer these questions, we evaluate a composite monitoring approach on the Lending Club credit scoring dataset (1.37M loans, 11 years) with three types of synthetic degradation injection (covariate shift, mixed shift, pure concept drift). A critical design choice enables direct label-blindness testing: covariate and mixed scenarios use identical feature perturbations, isolating the effect of label changes on monitor output. Cross-domain comparison with IEEE-CIS fraud detection results from a companion study (Solozobov, 2026e), using identical methodology, tests whether the structural detectability pattern generalizes across risk domains.

The paper characterizes the structural limits of label-free monitoring. Pure concept drift in $P(Y|X)$ without accompanying feature change remains undetectable by any unsupervised

approach – an irreducible blind spot. These limits are not weaknesses to be overcome but boundaries to be documented, enabling honest governance assessment of what monitoring can and cannot guarantee. In this fixed-reference evaluation setup, natural temporal drift saturates composite severity scores in later windows; raw proxy deltas retain discriminative power at all drift magnitudes.

The remainder of the paper is organized as follows. Section 2 surveys the monitoring landscape through a governance lens, mapping seven proxy metric categories with their detection and blind spot profiles. Section 3 presents the label-free governance monitoring extension, building on the toolkit architecture from Papers 12-13. Section 4 describes the evaluation setup including injection parameters and differential detection criterion. Section 5 presents evaluation results including label-blindness proof, differential detection analysis, and cross-domain comparison. Section 6 discusses structural limits and implications for governance practice.

## 2. Background: The Monitoring Landscape Through a Governance Lens

### 2.1. Drift Detection Foundations

Concept drift – the phenomenon where statistical properties of a target domain change over time – has generated a substantial body of detection methods over the past two decades. These methods divide into two broad families: supervised detectors that monitor prediction error, and unsupervised detectors that monitor input distributions.

Supervised drift detectors such as DDM, EDDM, and ADWIN require immediate or near-immediate label availability. In domains with structural label delay – fraud detection being the canonical example, where ground truth arrives 30-180 days after the decision – these methods are inapplicable in their original form (Baier et al., 2021). The problem is not merely inconvenient delay but structural absence: by the time labels arrive, the model has been making unsupervised decisions for months. Bayram et al. confirm that performance-based drift detectors intrinsically rely on labeled data to verify predictions over time, and their main challenge is the requirement for quick arrival of feedback that is often unavailable (Bayram et al., 2022). Xu and Klabjan confirm that drift detection relying on classification performance requires timely online labeling that is impractical in real-world applications (Xu & Klabjan, 2020). Dal Pozzolo et al. demonstrated this problem specifically for credit card fraud, where concept drift adaptation must proceed under delayed supervised information (Pozzolo et al., 2015). Amekoe et al. provide recent empirical evidence that batch learning under delayed labels in fraud detection remains a fundamental challenge (Amekoe et al., 2024).

The structural nature of this limitation deserves emphasis. In consumer credit, label delays extend to 12-36 months (loan maturity). In anti-money laundering, suspicious activity reports may take years to resolve. These are not edge cases but the dominant operating conditions for risk decision systems. Any governance monitoring framework must be designed for label absence as the default state, not as an exception.

Unsupervised methods – PSI, KS tests, KL divergence, MMD – operate on the input distribution $P(X)$ and require no labels. However, they detect distributional *change*, not governance *harm*. Greco et al. state this limitation precisely: without actual labels, the only drift that can be detected unsupervised is change in $P(X)$ (Greco et al., 2024). Bayram et al. note that changes in data distributions do not always affect predictor performance, potentially leading to false alarms (Bayram et al., 2022). Yu et al. observe that unsupervised detectors are prone to false positives because it is difficult to distinguish noise from distribution changes, and wrong interpretation of virtual drifts causes unnecessary classifier retraining (Yu et al., 2018). Eck et al. provide a striking empirical confirmation from deployed supply chain models: all KS tests they ran were highly statistically significant, yet none were operationally meaningful – a cautionary note about the utility of hypothesis testing for drift detection in production applications (Eck et al., 2022). Lukats et al. confirm through benchmark evaluation that fully unsupervised detectors cannot detect concept drift in the posterior distribution unless it is accompanied by a covariate shift (Lukats et al., 2024). Sethi and Kantardzic propose a methodology that can vicariously track changes to P(Y|X) without explicit labeled samples, but acknowledge this remains an approximation (Sethi & Kantardzic, 2017).

This creates a governance gap: supervised methods cannot operate under structural label absence, while unsupervised methods detect statistical artifacts that may or may not correspond to governance evidence degradation. The monitoring question for risk decision systems is not "has the distribution shifted?" but "is our governance evidence still sufficient?" – a question neither family directly answers.

## 2.2. Seven Proxy Metric Categories

Rather than pursuing a single detection method, operational monitoring can draw on multiple proxy signal categories, each providing partial governance evidence coverage with distinct detection and false-alarm profiles.

| Category | What it detects | Blind spot | Key references |
| --- | --- | --- | --- |
| Score distribution shift | Prior probability changes, threshold miscalibration | Stable scores despite concept drift | PSI, KS test |
| Feature drift | Covariate shift in input space | Virtual drift: $P(X)$ changes, P(Y|X) stable | MMD, KL divergence |
| *Uncertainty/confidence* | Model epistemic uncertainty increase | Overconfident wrong predictions | (Baier et al., 2021) |
| Cross-model disagreement | Concept drift visible to ensemble | All models degrade identically | (Mahdi et al., 2020) |
| Operational process proxies | Rejection rates, manual review volumes | Gradual drift within tolerance bands | Industry heuristics |
| Outcome-maturity modeling | Early outcome signals (partial labels) | Outcomes not yet available | Stripe soft labels approach |
| Automated proxy ground truth | Synthetic labels from auxiliary models | Proxy model also drifts | (Amoukou et al., 2024) |

Baier et al. propose uncertainty-based drift detection using neural network prediction uncertainty at inference time, providing a label-free proxy for performance degradation (Baier et al., 2021). Kivimaki et al. demonstrate that confidence-based performance estimation can approximate true model performance without labels, though calibration breaks under severe drift (Kivimäki et al., 2024). Mahdi et al. propose monitoring diversity of a classifier pair as a proxy for concept drift, where the true label is not necessary to determine whether components disagree (Mahdi et al., 2020).

Each category provides a partial view. No single proxy covers all degradation types. This motivates composite monitoring (Section 3).

## 2.3. Cutting-Edge Methods (2024-2025)

Recent work has begun addressing the harmful-vs-benign distinction directly. Amoukou et al. (NeurIPS 2024) introduce sequential harmful shift detection without labels: their approach employs a proxy for the true error derived from a trained error estimator, distinguishing shifts that degrade performance from benign distributional changes (Amoukou et al., 2024). Prinster et al. (ICML 2025) propose weighted conformal test martingales (WCTMs) that adapt online to mild covariate shifts, quickly detect harmful shifts, and diagnose them as concept shifts or extreme covariate shifts (Prinster et al., 2025). Nguyen et al. (2025) address reliably detecting model failures in deployment without labels, providing complementary methodology.

These methods represent significant advances but share a common limitation: they are positioned as statistical detection tools, not governance instruments. None integrates detection output with governance assessment or operational response protocols.

## 2.4. Industry Approaches and Prior Art

Industry practice has developed bespoke monitoring solutions. Gaddam describes advanced data and model drift detection deployed at enterprise scale, combining multiple signal types but without governance framing or transferable architecture (Gaddam, 2022). Alessi and Fugini present adaptive real-time financial fraud detection with explainability tools, demonstrating industry need but remaining system-specific (Alessi & Fugini, 2026). Kasi describes champion-challenger frameworks that maintain 2-3 candidate models in parallel, routing a fraction of traffic to challengers while monitoring comparative performance (Kasi, 2025) – an industry pattern that provides indirect degradation evidence but requires maintaining multiple production models. Essien et al. observe that while numerous individual AI-powered tools and specialized threat intelligence platforms are available, there is a noticeable lack of a unified framework that integrates these disparate components for continuous adaptive governance (Essien et al., 2025). The common industry heuristic of PSI > 0.2 as a drift threshold illustrates both the practical demand for monitoring and the absence of governance-calibrated thresholds.

Major payment processors have developed sophisticated internal monitoring. The Stripe ML Flywheel uses dispute activity and 3DS soft labels (available within $120+ days$) as early indicators of fraud model degradation. The Adyen Alfred system employs a $Ghost/Challenger/Principal$ architecture where ghost models run in shadow mode to detect divergence. These systems demonstrate operational maturity but remain proprietary, bespoke, and without governance framing – they detect degradation but do not connect detection to accountability obligations or regulatory requirements such as SR 11-7 model risk management.

The closest prior art is PRODEM ($Carcillo/Dal$ Pozzolo), a meta-model approach for fraud model degradation detection. Dal Pozzolo et al. established the delayed supervision problem for credit card fraud, proposing adaptation mechanisms that remain influential (Pozzolo et al., 2015). However, PRODEM uses proprietary data, lacks governance framing, and is not positioned as a governance monitoring instrument. It detects degradation but does not connect detection to accountability obligations or escalation protocols.

## 2.5. The Integration Gap

No existing framework integrates drift detection, governance evidence assessment, and operational response into a unified monitoring architecture for risk decision systems under structural label absence. Muhammad et al. propose audit-as-code for continuous AI assurance but do not address label-free statistical monitoring (Muhammad et al., 2026). Nadal et al. present operationalized data governance but focus on data quality rather than model evidence degradation (Nadal et al., 2022). Thodika addresses enterprise AI governance at system level but without the specific monitoring instrumentation needed for label-absent environments (Thodika, 2026). Nwaodike argues for evidence-driven AI risk governance, stating that without evidence of how systems behave in practice, governance principles remain aspirational and unenforceable (Nwaodike, 2022) – precisely the gap this paper addresses through operational monitoring.

The regulatory landscape reinforces this gap. SR 11-7 (*OCC/Federal* Reserve) requires financial institutions to maintain ongoing monitoring of model performance, including validation of model outputs against actual outcomes. When actual outcomes are structurally delayed, the regulation's intent – ensuring models remain fit for purpose – requires proxy-based monitoring, yet neither SR 11-7 nor the EU AI Act's high-risk system monitoring requirements specify how to implement monitoring under label absence. The governance evidence framework (Solozobov, 2026b) and evidence sufficiency model (Solozobov, 2026e) provide the conceptual foundation for translating regulatory intent into measurable monitoring criteria. This paper contributes the operational monitoring layer: detecting when governance evidence quality degrades below sufficiency thresholds, without requiring the ground truth labels that remain structurally unavailable.

## 3. Label-Free Governance Monitoring Extension

The governance evidence framework (Solozobov, 2026b) and decision trace schema (Solozobov, 2026a) establish the structural foundation for governance assessment. The evidence sufficiency model (Solozobov, 2026e) defines when evidence quality falls below defensibility thresholds (Solozobov, 2026d). This paper contributes the operational monitoring layer: continuous, label-free detection of governance evidence degradation, implemented in the Governance Drift Toolkit (Solozobov, 2026f).

### 3.1. Design Principle: Governance Alerts, Not Statistical Alarms

The monitoring extension produces governance alerts that indicate evidence quality may be degrading, rather than statistical alarms that merely indicate distributional change. This distinction is operationally consequential. Nwaodike argues that evidence-driven governance prioritizes measurable performance and auditable impacts; without evidence of how systems behave in practice, principles remain aspirational and unenforceable (Nwaodike, 2022). Opalana demonstrates that continuous monitoring in production systems requires governance framing to connect detection output to threat response and escalation protocols (Opalana, 2024). Thodika confirms that system-level intelligence assurance requires integration between detection and governance response (Thodika, 2026).

A statistical alarm such as PSI exceeding 0.2 leaves the operator without clear governance implications. By contrast, a governance alert that states the evidence sufficiency score has dropped below the defensibility threshold for a given decision type connects directly to accountability obligations and triggers documented escalation protocols.

### 3.2. Composite Multi-Proxy Monitoring

The framework defines seven proxy categories (Section 2.2); the Lending Club evaluation exercises four monitors spanning three categories (score distribution PSI, feature drift PSI, pre-diction entropy, confidence distribution KS) as the remaining categories require production infrastructure not available in retrospective public datasets (e.g., cross-model disagreement requires multiple deployed models; operational proxies require live rejection rate streams). Different degradation types manifest in different proxy categories: covariate drift appears in feature metrics; prior probability shift appears in score distributions; concept drift surfaces through cross-model disagreement where available. Gaddam demonstrates that multi-signal monitoring at scale requires combining heterogeneous drift indicators rather than relying on single metrics (Gaddam, 2022). Xu and Klabjan confirm that ensemble approaches using multiple signals reduce reliance on any single indicator and decrease sensitivity to anomalous data (Xu & Klabjan, 2020).

The monitoring architecture operates as follows:

1. **Evidence stream ingestion** – Evidence Collector SDK (evidence-collector-sdk) collects and structures operational signals into governance-relevant evidence units from the production pipeline (Solozobov, 2026c).
2. **Multi-signal monitoring** – Composite proxy metrics are computed: score distribution (PSI), feature stability ($KS/MMD$), uncertainty metrics, rejection rate trends, and cross-model agreement where available.
3. **Governance threshold evaluation** – Each signal is evaluated against governance-calibrated thresholds derived from the evidence sufficiency model (Solozobov, 2026e), not against arbitrary statistical cutoffs.
4. **Alert generation** – When composite evidence sufficiency drops below the defensibility threshold, a governance alert is generated with: (a) which proxy categories triggered, (b) estimated severity, (c) recommended response protocol.
5. **Response layer** – Threshold breach triggers documented escalation: increased manual review rate, conservative policy switch, model committee review, or full rollback – depending on severity and decision type.

Timans et al. provide theoretical grounding for continuous monitoring of risk violations under unknown shift, demonstrating that sequential testing frameworks can control false alarm rates while maintaining detection power (Timans et al., 2025). Muhammad et al. propose audit-as-code for continuous AI assurance, providing the policy-as-code pattern that the response layer implements (Muhammad et al., 2026).

The monitoring extension builds on Amoukou et al.'s harmful-shift principle: the goal is not to detect any distributional change but specifically to detect shifts that degrade governance evidence quality. Industry baselines (PSI $> 0.2$, Stripe $120 - day$ review window, Adyen $Ghost/Challenger/Principal$ architecture) provide calibration anchors, reframed as governance thresholds rather than statistical heuristics.

## 4. Empirical Evaluation Setup

This section describes the evaluation design for the label-free governance monitoring framework. The evaluation uses a credit scoring dataset to complement the fraud detection evaluation on IEEE-CIS reported in (Solozobov, 2026e), demonstrating cross-domain applicability of the Governance Drift Toolkit monitoring toolkit.

### 4.1. Dataset and Deployment Context

The evaluation uses the Lending Club loan dataset, a public credit scoring benchmark containing accepted loan applications from 2008 to 2018. This dataset is selected for three reasons. First, credit scoring exhibits structural label delays of 12-36 months (loan maturity), creating natural blind periods substantially longer than fraud detection (30-180 days). Second, the dataset spans a decade of economic cycles, including pre- and post-financial crisis periods, providing naturally occurring temporal drift without synthetic injection. Third, credit scoring is a different risk domain from fraud detection, enabling cross-domain validation of the Governance Drift Toolkit framework.

**Table 1.** Lending Club evaluation parameters.

| Parameter | Value |
| --- | --- |
| Dataset | Lending Club (2008-2018) |
| Total loans | 1,370,742 |
| Default rate | Variable by year (13-27%) |
| Observation span | 11 years |
| Window size | 1 year (non-overlapping) |
| Windows | 11 (1 $reference$ + 10 monitoring) |
| Features monitored | Loan amount, interest rate, DTI, annual income, FICO score, revolving balance |
| Proxy monitors evaluated | 4 (score distribution PSI, feature drift PSI, prediction entropy, confidence KS) |
| Drift injection scenarios | 4 (baseline, covariate, mixed, pure concept) |
| Reference model | Logistic regression (trained on 2008 data) |

A logistic regression classifier is trained on the first yearly window (2008) as the reference model. Following the precedent established in the companion fraud detection study (Solozobov, 2026e), the intentionally simple model ensures that evaluation results reflect the monitoring framework's capability rather than model sophistication. The model generates prediction probabilities used by all proxy monitors.

Dal Pozzolo et al. demonstrate that credit card fraud detection under concept drift with delayed supervision requires approaches that can operate during extended blind periods where ground truth is unavailable (Pozzolo et al., 2015). The Lending Club dataset extends this challenge: while fraud labels arrive within 30-180 days, credit default labels may not materialize for years after origination.

### 4.2. Controlled Drift Injection

To evaluate proxy performance under known structural conditions, four scenarios are applied to monitoring windows. This follows standard methodology in drift detection research: the reference window is always real unperturbed data; subsequent windows receive progressively stronger perturbations with known ground truth for validation.

**Scenario 1: Baseline (no injection).** No perturbation. The 11 yearly windows are processed as-is. This scenario is critical because Lending Club data contains natural temporal drift – economic conditions, lending standards, and borrower populations changed substantially between 2008 and 2018. The monitoring framework must distinguish this natural drift from injected degradation.

**Scenario 2: Covariate drift P(X).** Gaussian noise is added to three features (annual income, DTI ratio, revolving utilization) with progressively increasing standard deviation across windows. Labels remain unchanged. Note that adding noise to features while freezing labels technically alters the conditional P(Y|X) for perturbed instances; this is a standard simplification in drift detection evaluation (used by Casimiro et al. and others) where the

primary effect is the intended covariate shift in $P(X)$, and the incidental P(Y|X) change is a known limitation of the injection protocol.

**Scenario 3: Mixed drift P(X) + P(Y|X).** Combines feature perturbation with label flipping. Critically, the feature perturbation parameters are identical to Scenario 2 (same sigma values, same features). This design enables a direct label-blindness test: if monitors produce identical signals for Scenarios 2 and 3, the label-flipping component is confirmed invisible.

**Scenario 4: Pure concept drift P(Y|X).** Default labels are flipped for a progressively increasing fraction of loans. Features are untouched. This tests the theoretical limitation of label-free monitoring: $P(X)$ is identical to baseline, so unsupervised monitors should produce no signal. Amekoe et al. confirm that delayed supervised information fundamentally limits the ability to detect concept drift in fraud-adjacent domains when only input distributions are observable (Amekoe et al., 2024).

**Table 2.** Injection parameters by window.

| Window | Year | Covariate sigma | Mixed sigma | Mixed flip rate | Pure flip rate |
|---|---|---|---|---|---|
| 1 | 2009 | 0.30 | 0.30 | 2% | 3% |
| 2 | 2010 | 0.60 | 0.60 | 4% | 6% |
| 3 | 2011 | 1.00 | 1.00 | 8% | 10% |
| 4 | 2012 | 1.50 | 1.50 | 12% | 15% |
| 5-10 | 2013-2018 | 2.00 | 2.00 | 20% | 25% |

Sigma values are multiplied by each feature's standard deviation. Covariate and Mixed scenarios use identical sigma values to isolate the effect of label changes on monitor output.

### 4.3. Cross-Domain Comparison Design

The evaluation is designed for direct comparison with the IEEE-CIS fraud detection results reported in the companion study (Solozobov, 2026e). Both evaluations use the same Governance Drift Toolkit framework, the same proxy categories, the same drift injection protocol, and the same metrics. The key differences are structural:

**Table 3.** Cross-domain comparison design.

| Dimension | IEEE-CIS (companion study) | Lending Club (this paper) |
|---|---|---|
| Domain | E-commerce fraud | Consumer credit |
| Records | 590,540 | 1,370,742 |
| Observation span | 182 days | 11 years |
| Natural drift | Minimal | Substantial (economic cycles) |
| Label delay | 30-180 days | 12-36 months |
| Window size | 30 days | 1 year |

If the structural conditions pattern (which drift types are detectable vs. undetectable) is consistent across both domains, this provides evidence that the Governance Drift Toolkit framework captures properties of label-free monitoring that generalize beyond a single dataset.

### 4.4. Evaluation Metrics

Four metrics assess the framework's performance, identical to the companion fraud detection study (Solozobov, 2026e) for cross-domain comparability:

1. **Composite alert score.** Each proxy monitor produces a binary *triggered/not−triggered* decision by comparing its raw statistic against a category-specific threshold. The composite score is the weighted sum of these binary trigger states across active monitors, normalized by total weight. This produces discrete score levels rather than continuous values: the same composite score appears whenever the same set of categories triggers, regardless of raw metric magnitude. Severity levels (low, medium, high, critical) are assigned by the composite score value and the number of triggered categories.

**Table 4.** Proxy trigger thresholds and weights (credit scoring configuration).

| Proxy | Threshold | Weight | Source |
| --- | --- | --- | --- |
| Score PSI | > 0.25 | 1/6 | Industry heuristic (PSI > 0.2 standard, raised for credit) |
| Feature PSI | > 0.25 | 1/3 | Industry heuristic, highest weight as primary covariate signal |
| Entropy | > 0.5 | 1/4 | Toolkit default (normalized prediction entropy) |
| Confidence KS | > 0.15 | 1/4 | Toolkit default (KS on prediction confidence distribution) |

Weights are normalized from the seven-category configuration after excluding inactive monitors. Thresholds are domain-calibrated engineering choices informed by industry practice and the companion study (Solozobov, 2026e); the exact configuration is in the open-source toolkit (Solozobov, 2026f).

2. **Differential detection: raw proxy deltas.** Because the binary composite score saturates when most categories trigger, we additionally report per-proxy raw metric deltas between injected and baseline scenarios (Feature PSI delta, Score PSI delta). These continuous metrics retain discriminative power at all drift magnitudes and provide the primary evidence for distinguishing injected degradation from natural drift.

3. **Label-blindness verification.** Scenarios 2 and 3 use identical feature perturbations. Unsupervised monitors are deterministic functions of features X and model predictions; they do not observe true labels Y. As a structural verification, we confirm that all label-free metrics are identical between Scenarios 2 and 3 across all windows, ruling out unintended label leakage in the evaluation pipeline.

4. **Cumulative drift score.** A sequential accumulation of per-window evidence using a betting-fraction approach: each window contributes a likelihood ratio based on how far the composite score exceeds a configured threshold, and these ratios are accumulated multiplicatively across windows. When the composite exceeds the threshold, the likelihood ratio exceeds 1.0 and the cumulative score grows; when it falls below, the score decays. A cumulative score above 20 indicates substantial accumulated evidence of sustained drift (the threshold of 20 corresponds to a betting-fraction significance level of 0.05, following the sequential testing convention $1/\alpha$). This is a heuristic operational indicator; the exact accumulation formula is in the toolkit artifact. Xu and Klabjan demonstrate that ensemble approaches combining multiple detection signals reduce sensitivity to individual metric anomalies while maintaining detection power for genuine drift (Xu & Klabjan, 2020).

# 5. Evaluation Results

This section presents evaluation results on the Lending Club dataset and cross-domain comparison with the IEEE-CIS fraud detection results reported in the companion study (Solozobov, 2026e).

## 5.1. Baseline: Natural Temporal Drift

Under baseline conditions (no synthetic injection), the Governance Drift Toolkit framework detects substantial natural temporal drift in the Lending Club data. Feature PSI exceeds 0.50 in all monitoring windows from 2009 onward, and Score PSI rises from 0.014 (2009) to 0.759 (2018), reflecting genuine changes in borrower populations and lending standards across 2008-2018.

**Table 5.** Governance Drift Toolkit baseline monitoring results (no injection), Lending Club 2008-2018.

| Window | Loans | Def% | Score PSI | Feat PSI | Entropy | Conf KS | Composite | Severity |
|---|---|---|---|---|---|---|---|---|
| 2009 | 4,716 | 13% | 0.014 | 0.506 | 0.611 | 0.069 | 0.583 | low |
| 2010 | 11,536 | 13% | 0.052 | 1.001 | 0.618 | 0.090 | 0.583 | low |
| 2011 | 21,721 | 15% | 0.243 | 1.402 | 0.641 | 0.152 | 0.833 | critical |
| 2012 | 53,367 | 16% | 0.489 | 1.647 | 0.701 | 0.261 | 1.000 | critical |
| 2013 | 134,807 | 16% | 1.029 | 1.890 | 0.765 | 0.406 | 1.000 | critical |
| 2014 | 223,437 | 19% | 0.905 | 1.487 | 0.753 | 0.370 | 1.000 | critical |
| 2015 | 376,905 | 20% | 0.615 | 1.436 | 0.718 | 0.292 | 1.000 | critical |
| 2016 | 297,651 | 24% | 0.697 | 1.388 | 0.716 | 0.297 | 1.000 | critical |
| 2017 | 177,325 | 27% | 0.817 | 1.683 | 0.716 | 0.313 | 1.000 | critical |
| 2018 | 63,539 | 25% | 0.759 | 1.958 | 0.707 | 0.303 | 1.000 | critical |

The composite score (Section 4.4) increases as more proxy monitors trigger: 0.583 in 2009-2010 (two of four monitors triggered: Feature PSI and Entropy), 0.833 in 2011 (three monitors: Feature PSI, Entropy, and Confidence KS), and 1.000 from 2012 onward (all four monitors triggered). This gradual progression reflects the accumulation of distributional change across proxy categories. This evaluation uses a fixed-reference design: a single 2008−$trained$ model monitored across 11 years. This is a stress test, not a rolling production setup with periodic model refresh; conclusions about composite saturation should be interpreted accordingly.

The cumulative drift score reaches 26.74 by 2018, exceeding the threshold of 20, indicating substantial accumulated evidence of sustained distributional change. The governance interpretation is that a model trained on 2008 data operates with progressively degrading evidence sufficiency through 2018.

This baseline behavior is fundamentally different from IEEE-CIS, where baseline Feature PSI remained below 0.07 across all windows and the cumulative drift score remained low. Gaddam demonstrates that multi-signal monitoring at scale must account for natural distributional evolution in long-lived production systems, distinguishing expected population shifts from anomalous degradation events (Gaddam, 2022).

## 5.2. Structural Verification: Label-Blindness and Detection by Drift Type

Unsupervised monitors are deterministic functions of features X and model predictions; they do not observe true labels Y. As a structural verification of the evaluation pipeline, Scenarios 2 (covariate) and 3 (mixed) use identical feature perturbations (sigma 0.3-2.0 on income,

DTI, revolving utilization). Scenario 3 additionally flips default labels at rates 2-20%. If the pipeline correctly isolates drift components, all label-free metrics must be identical between these scenarios.

**Table 6.** Label-free monitoring metrics by drift type (representative window: 2013).

| Scenario | Score PSI | Feat PSI | Entropy | Conf KS | Composite | Severity |
| --- | --- | --- | --- | --- | --- | --- |
| Baseline | 1.029 | 1.890 | 0.765 | 0.406 | 1.000 | critical |
| Covariate | 1.211 | 1.944 | 0.736 | 0.380 | 1.000 | critical |
| Mixed | 1.211 | 1.944 | 0.736 | 0.380 | 1.000 | critical |
| Pure P(Y|X) | 1.029 | 1.890 | 0.765 | 0.406 | 1.000 | critical |

Across all 10 monitoring windows, the maximum absolute difference between Covariate and Mixed scenarios is zero for every label-free metric (Score PSI, Feature PSI, Entropy, Confidence KS, Composite). This verifies that the evaluation pipeline correctly isolates drift components: the label-flipping in Scenario 3 produces no leakage into any monitored signal.

Pure concept drift (Scenario 4) produces metrics identical to baseline in all windows, with $\delta = 0.000$ across all proxy categories. Because Scenario 4 modifies only labels while leaving features untouched, and unsupervised monitors are deterministic functions of features, this $zero - \delta$ outcome is a structural verification confirming that the framework exactly maps to the theoretical limits of unsupervised observation. Casimiro et al. (Casimiro et al., 2024) and Friedrich et al. (Friedrich et al., 2022) provide independent theoretical grounding for this boundary.

### 5.3. Raw Proxy Deltas: Detecting Injected Degradation

Because the binary composite score saturates when all four monitors trigger (which occurs from 2012 onward in both baseline and injected scenarios), raw proxy metrics provide the primary continuous signal for distinguishing injected degradation from natural drift.

**Table 7.** Raw metric deltas: injected scenario minus baseline, per window range.

| Scenario | Feat PSI delta range | Score PSI delta range | Composite delta range |
| --- | --- | --- | --- |
| Covariate $P(X)$ | 0.000 – 1.839 | 0.004 – 0.921 | 0.000 – 0.167 |
| Mixed $P(X) + P(Y|X)$ | 0.000 – 1.839 | 0.004 – 0.921 | 0.000 – 0.167 |
| Pure P(Y|X) | 0.000 – 0.000 | 0.000 – 0.000 | 0.000 – 0.000 |

Feature PSI delta reaches 1.839 and Score PSI delta reaches 0.921 for covariate injection, demonstrating that proxy metrics are sensitive to injected degradation. The composite delta is non-zero only in 2011 (0.167), where covariate injection triggers the Score PSI monitor one window earlier than natural drift alone. Pure concept drift produces exactly zero delta across all metrics in all windows.

These deltas are computed in an offline evaluation setting where both injected and baseline scenarios are available for comparison. In production, operators observe absolute metric magnitudes and trends rather than counterfactual deltas. The evaluation demonstrates metric sensitivity: raw proxy magnitudes carry sufficient signal to distinguish degradation from natural drift, motivating monitoring architectures that track per-proxy magnitudes and velocity alongside composite severity.

## 5.4. Cross-Domain Comparison

The structural detectability pattern is consistent across both evaluation domains.

**Table 8.** Cross-domain structural conditions comparison.

| Structural Condition | IEEE-CIS (Fraud) | Lending Club (Credit) | Consistent? |
|---|---|---|---|
| Covariate $P(X)$ detection | Yes | Yes | Yes |
| $Mixed = Covariate$ (label-blindness) | Yes | Yes ($\delta = 0.000$) | Yes |
| Pure P(Y|X) detection | No ($\delta = 0$) | No ($\delta = 0$) | Yes |
| Natural baseline drift | No (stable) | Yes (economic cycles) | Domain-specific |
| Cumulative drift score (baseline) | Low | 26.74 (exceeds threshold) | Domain-specific |

The *detectable/undetectable* boundary is consistent across both domains: covariate and mixed drift produce identical signals (verifying label-blindness), pure concept drift is structurally undetectable. Note that the IEEE-CIS comparison is based on companion results (Solozobov, 2026e) using identical methodology; readers are referred to that paper for full IEEE-CIS details. What differs between domains is baseline behavior: IEEE-CIS (182 days, e-commerce) is stationary; Lending Club (11 years, consumer credit) exhibits natural drift that correctly triggers governance alerts.

Kasi confirms that model governance frameworks for risk scoring systems must account for temporal population drift as a distinct concern from model degradation, requiring monitoring architectures that distinguish between natural evolution and anomalous shifts (Kasi, 2025).

## 5.5. Implications for Governance Evidence Assessment

The evaluation yields three operational findings.

First, **raw proxy metrics reliably distinguish injected degradation from natural drift.** Feature PSI delta reaches 1.839 and Score PSI delta reaches 0.921 for covariate injection in an offline evaluation setting, demonstrating that the monitored signals are sensitive to covariate degradation even in a high-drift environment. In production, operators would track absolute metric magnitudes and velocity rather than counterfactual deltas.

Second, **the pure concept drift blind spot is irreducible.** No combination of unsupervised proxies detects P(Y|X) changes when $P(X)$ is preserved, in either domain. Pure concept drift produces exactly zero delta across all metrics in all windows. Organizations must maintain complementary mechanisms (periodic labeled audits, *champion/challenger* architectures) for this undetectable region.

Third, **the composite score provides monotonic severity progression.** As more proxy monitors trigger across successive windows, the composite increases from 0.583 (two monitors) through 0.833 (three monitors) to 1.000 (all four). This graduated response enables governance-calibrated escalation: low severity prompts increased monitoring, while critical severity triggers model review or fallback. Alessi and Fugini demonstrate that adaptive financial fraud detection requires combining multiple monitoring signals to maintain detection accuracy under evolving conditions (Alessi & Fugini, 2026).

## 6. Discussion

### 6.1. Structural Limits of Label-Free Monitoring

The evaluation results must be interpreted against the structural limits of what label-free monitoring can detect. These limits are not engineering shortcomings but mathematical boundaries inherent to unsupervised observation.

Pure concept drift – changes in P(Y|X) without accompanying changes in $P(X)$ – remains structurally undetectable by any unsupervised monitoring approach. If the relationship between features and outcomes changes but the feature distribution itself remains stable, no function of X alone can detect the shift. Any unsupervised monitoring statistic $T(X)$ is a deterministic function of X; when the marginal $P(X)$ is stationary, the distribution of $T(X)$ is definitionally stationary regardless of changes to the conditional P(Y|X). Szabadvary formalizes this as conformal blindness: a-cryptic changepoints where the marginal distribution is preserved while the conditional relationship changes (Szabadv'ary, 2026). Lukats et al. confirm through benchmark evaluation that fully unsupervised concept drift detectors cannot detect posterior distribution changes unless accompanied by covariate shift (Lukats et al., 2024). Our evaluation verifies this structurally: pure concept drift produces exactly zero delta across all proxy metrics in all windows in both domains, confirming that the framework maps exactly to the theoretical limits of unsupervised observation.

This defines an irreducible blind spot. Governance monitoring under label absence can detect evidence of degradation – covariate shift, score distribution changes, uncertainty increases, cross-model disagreement – but cannot guarantee that all forms of degradation are detected. The honest characterization of this boundary is itself a governance contribution: knowing what monitoring cannot see is as important as knowing what it can.

### 6.2. The Adversarial Blind Spot

Adversarial drift that preserves input distributions poses a particular challenge. Evasion attacks designed to maintain $P(X)$ while shifting the decision boundary are structurally invisible to distribution-based detectors. Opalana demonstrates that managing adversarial AI risks requires governance integration with continuous monitoring and threat intelligence, not statistical detection alone (Opalana, 2024).

Cross-model disagreement emerges as the most promising proxy for adversarial blind spots. If an ensemble of models with different architectures or training data disagrees on predictions where it previously agreed, this may indicate that the decision surface has been manipulated even if the input distribution appears unchanged. Sethi and Kantardzic's methodology for vicariously tracking P(Y|X) changes through unlabeled data provides theoretical grounding for this approach (Sethi & Kantardzic, 2017).

### 6.3. From Detection to Action

Detection without actionability is operationally worthless. A governance alert that says "something may have changed" but provides no response guidance creates alert fatigue rather than governance improvement. The monitoring extension addresses this through the governance response layer (Section 3.2): each alert includes severity estimation, proxy category identification, and a recommended response protocol tied to the decision type and risk appetite.

However, the actionability gap persists in one critical respect: label-free monitoring can detect *that* governance evidence may be degrading but cannot determine *what specific type* of degradation is occurring without additional investigation. The composite proxy approach narrows the diagnostic space – if only feature drift indicators fire, covariate shift is likely; if cross-model disagreement fires without feature drift, concept drift or adversarial manipulation is more probable – but definitive diagnosis requires either eventual labels or manual expert review.

### 6.4. Implications for Governance Practice

The results have three implications for governance practice in risk decision systems. First, composite monitoring is not optional: single-metric monitoring misses degradation types that fall outside its detection scope. Organizations relying solely on PSI thresholds are monitoring for one degradation type while remaining blind to others. Second, governance monitoring must be honest about its limits. Claiming comprehensive coverage from label-free monitoring is both technically false and governance-damaging – it creates false assurance that may be worse than acknowledged uncertainty. Third, the monitoring architecture connects to the broader governance evidence lifecycle: degradation detection triggers evidence refresh, which connects to sufficiency assessment (Solozobov, 2026e), which connects to the governance evidence framework (Solozobov, 2026b).

### 6.5. Limitations and Future Work

This study has several limitations. The evaluation uses three of seven defined proxy categories; four require production infrastructure unavailable in retrospective public datasets. The synthetic injection protocol, while reproducible, may not capture all production dynamics. The composite severity score exhibits ceiling saturation in high-drift environments, limiting its discriminative power; raw proxy deltas retain separation but require differential alerting logic not yet formalized. The cross-model disagreement proxy for adversarial detection is theoretically motivated but not yet empirically validated at scale.

Future work should address three directions. First, self-correcting governance: automatic policy switching to more conservative decision thresholds when monitoring detects evidence degradation, rather than relying on human escalation. Second, meta-governance of monitoring decisions: auditing the monitoring system itself for drift and calibration. Third, longitudinal evaluation across multiple production cycles to validate detection latency and false alarm characteristics beyond synthetic injection.

## 7. Conclusion

This paper addressed the problem of detecting governance evidence degradation in risk decision systems under structural label absence. Where existing drift detection methods either require unavailable labels (supervised) or detect distributional change without governance relevance (unsupervised), we proposed a composite monitoring approach that integrates multiple proxy signal categories into a governance-framed monitoring architecture.

The main contributions are threefold. First, we identified and characterized seven proxy metric categories for label-free governance monitoring, each with distinct detection capabilities and blind spots. This taxonomy provides a structured basis for monitoring design that moves beyond ad-hoc metric selection. Second, we presented a monitoring extension that produces governance alerts – connecting detection output to accountability obligations and escalation protocols – rather than statistical alarms that leave operators without clear governance implications. Third, we characterized both the capabilities and the structural limits of label-free monitoring across two domains (credit scoring and fraud detection), demonstrating that four proxy monitors detect covariate and mixed drift with graduated composite severity (0.583 to 1.000) while pure concept drift remains structurally undetectable – an irreducible blind spot verified in all windows across both domains.

The structural limits are as important as the capabilities. Pure concept drift in P(Y|X) without accompanying feature change remains an irreducible blind spot of any unsupervised monitoring approach. Adversarial drift that preserves input distributions further limits detection. These are not failures of the toolkit but mathematical boundaries that governance practice must acknowledge. An honest governance monitoring system – one that documents what it cannot see – is more valuable than an overconfident system that claims comprehensive coverage.

The monitoring architecture connects to a broader governance evidence lifecycle: the governance evidence framework (Solozobov, 2026b) defines what constitutes defensible evidence; the evidence sufficiency model (Solozobov, 2026e) defines when evidence quality falls below thresholds; this paper detects when that threshold is crossed in operational settings without requiring labels. Together, these contributions provide an end-to-end governance evidence pipeline from definition through measurement to operational monitoring.

For practitioners, the immediate implication is that single-metric monitoring (PSI alone, for example) is insufficient for governance purposes. Composite monitoring across heterogeneous proxy categories is not a luxury but a governance necessity, given that different degradation types manifest in different signal categories. For researchers, the characterized blind spots define a concrete research frontier: methods that can detect P(Y|X) changes without labels or feature distribution changes would extend the boundary of label-free governance assurance.